\documentclass[aps, preprintnumbers, tightenlines, nofootinbib, twocolumn]{revtex4}

\usepackage{amssymb,amsmath,amsfonts,amsbsy,epsfig,wrapfig,cancel,graphicx,natbib}

\newcommand{\eq}[2]{\begin{equation}\label{#1}#2 \end{equation}}

\usepackage{xcolor}

\newcommand{\w}{\widetilde}

\newcommand{\calZ}{{\cal Z}}

\newcommand{\roughly}[1]{\mathrel{\raise.3ex\hbox{$#1$\kern-0.85em\lower1ex\hbox{$\sim$}}}}

\newbox\charbox
\newbox\slabox
\def\slsh#1{{      
        \setbox\charbox=\hbox{$#1$}
        \setbox\slabox=\hbox{$/$}
        \dimen\charbox=\ht\slabox
        \advance\dimen\charbox by -\dp\slabox
        \advance\dimen\charbox by -\ht\charbox
        \advance\dimen\charbox by \dp\charbox
        \divide\dimen\charbox by 2
        \raise-\dimen\charbox\hbox to \wd\charbox{\hss/\hss}
        \llap{$#1$}
}}

\def\nn{\nonumber}

\def\bea{\begin{eqnarray}}
\def\eea{\end{eqnarray}}
\def\be{\begin{equation}}
\def\ee{\end{equation}}

\def\bea{\begin{eqnarray}}
\def\eea{\end{eqnarray}}
\def\be{\begin{equation}}
\def\ee{\end{equation}}
\def\ba{\begin{array}}
\def\ea{\end{array}}

\def\nn{\nonumber}

\begin{document}
\author{Yaneer Bar-Yam$^{1}$}
\email{yaneer@necsi.edu}
\author{Subodh P. Patil$^{2}$\footnote{Corresponding author}}
\email{patil@nbi.ku.dk}
\affiliation{1) New England Complex Systems Institute,\\ 277 Broadway, Cambridge MA 02139, U.S.A\\}
\affiliation{2) Niels Bohr International Academy and Discovery Center,\\ Niels Bohr Institute, Blegdamsvej 17,\\ Copenhagen, DK 2100, Denmark\\}
\date{\today}

\title{Renormalization of Sparse Disorder in the Ising Model}

\begin{abstract}
We consider the renormalization of quenched bond disorder in the Ising model in the limit that it is sparse -- highly localized and vanishing in the thermodynamic limit. We begin in 1D with arbitrary disorder assigned to a finite number of bonds and study how the system renormalizes, finding non-trivial fixed point structure for any given bond with a separatrix at zero bond strength, equivalent to inserting a break in the chain. Either side of this critical line, renormalization group (RG) trajectories flow towards one of two attractors on which the disordered bonds settle onto ferromagnetic or anti-ferromagnetic (AF) couplings of equal and opposite magnitude. Bonds that settle on an AF attractor are equivalent to inserting a twist in the chain at the location of the bond, implying a multi-kink ground state solution. Qualitatively different behavior emerges at the RG step when bonds start to coalesce, with the chain `untwisting' whenever two AF bonds coalesce. Our findings generalize to higher dimensions for codimension one defects that are sparse from the perspective of the orthogonal complement lattice. In 2D, the $\mathbb Z_2$ symmetry of the model has an IR manifestation where one can construct field strengths and Wilson loops (which characterize frustration) from fundamental plaquette variables. In the non-sparse limit where the disorder parameter is drawn from an arbitrary but homogeneously assigned probability distribution function, we recover previously found fixed distributions as special cases, but find only the trivial paramagnetic distribution to be an attractor. For non-homogeneously assigned disorder that is sufficiently dilute however, we find the Edwards-Anderson model with equal probability $\pm J$ bonds to also be an IR attractor. 
\end{abstract}

\maketitle
\tableofcontents
\section{Introductory remarks}

The study of structural variations or quenched disorder in spin systems is of intrinsic interest due to the wide ranging relevance of spin models in representing collective behavior across diverse physical systems \cite{spin glass}. For example, in complex networks one often finds the existence of pre-specified small scale, localized structures and one would like to study their effects, if any, on the behavior of an interacting system at larger scales \cite{CN}. 

One might also want to study smaller scale systems where the effects of individual, or small collections of defects are the object of interest, as is the case in a variety of nano-technological and biological systems where heterogeneous structures are present. Given the ubiquitousness of spin models in describing the state space of various computational \cite{neural networks}, biological \cite{neurophysics} and chemical systems \cite{protein folding}, we find ample motivation to study the effects of quenched bond disorder in the Ising model that is \textit{sparse} -- that is, disorder that is highly localized and vanishing in the thermodynamic limit (though not necessarily a weak perturbation), and examine its behavior under renormalization. 

When disorder is non-sparse and statistical in nature, approximations such as the replica trick \cite{replica} have rendered the study of certain idealized systems tractable -- the Edwards-Anderson, Sherrington-Kirkpatrick and Sachdev-Ye-Kitaev models being prototypical examples \cite{EA, SK, SY, Kitaev} (the latter only being solvable in the limit of large spins). However, the case of highly localized, arbitrarily assigned disorder does not appear to have warranted as much attention as systems where the disorder at any particular location is taken to be a random variable drawn from a given probability distribution. In what follows, we consider quenched disorder in a spin system with vanishing external field, restricting ourselves to questions concerning scaling properties that are also tractable in the limiting case of sparse disorder. We define sparse disorder as arbitrarily assigned disorder localized to a finite number of bonds. Or, more generally, to a vanishing density in the thermodynamic limit. We compare to the non-sparse limit where appropriate. 

\subsection{Outline and summary of results}
We begin by considering a 1D Ising spin chain with vanishing external field, implementing renormalization group (RG) transformations through decimation. In the case of a single bond with arbitrary initial strength, parametrized as
$\beta_0(1 + g_0)$ with $g_0$ the (bare) bond disorder paremeter, we demonstrate a non-trivial fixed point structure that persists in the non-sparse limit. We find that any initial perturbation, however strong, rapidly flows towards one of two attractors on which the disorder parameter settles on values corresponding to a ferromagnetic/ anti-ferromagnetic coupling for an initially positive/ negative perturbation separated by a critical line at vanishing coupling ($g = -1$). 

In the absence of an external field, inserting an anti-ferromagnetic (AF) bond of equal and opposite strength into an otherwise ferromagnetic (F) spin chain with free boundary conditions is equivalent to flipping all spins on one side of the bond, that is, inserting a twist in the spin chain at the location of the bond. Thus for free boundary conditions the ground state in the presence of single AF bond disorder is a kink. This is sometimes referred to casually as a domain wall, by projection from a two dimensional defect in three spatial dimensions that results from the $\mathbb Z_2$ degeneracy of the ground state. For a spin chain with periodic boundary conditions, single AF bond disorder results in frustration which is topological (rather than geometrical) in nature arising as it does from the boundary conditions. 
  
Since we consider only nearest neighbor interactions, locality implies that two or more initially perturbed bonds renormalize independently until the RG step where any two coalesce. The configuration of the spin chain before any two bonds coalesce corresponds to a multi-kink ground state in the case of free boundary conditions. Upon coalescence, we find that any two bonds which have both settled on, or are close to the AF attractor will have `untwisted', whereas F bonds coalescing with AF bonds remain twisted. Thus small scale or mesoscopic disorder averages out at long wavelengths up to an overall twist that depends on the index 
\eq{wdef}{\mathcal W := (-1)^{N_0},} 
where $N_0$ counts the number of negative bare couplings (i.e. with individual  disorder parameters $\{g_0\} < -1$, no matter their magnitude), the non-vanishing of which would correspond to the presence of frustration for a periodic spin chain. Therefore in 1D, the only disorder that is relevant at long wavelengths in the absence of an external magnetic field is that which would correspond to frustration in the case of periodic boundary conditions. 

When considering the non-sparse limit in 1D, we recover the findings of Grinstein et al. \cite{Grinstein} as a special case, showing a restricted class of (spatially uniform) probability distribution functions (PDFs) for the disorder parameter to be \textit{fixed distributions} under renormalization. These solutions correspond to F, AF and paramagnetic phases in addition to one corresponding to a spin glass phase. However we find that with the exception of the (trivial) paramagnetic fixed distribution, the F, AF and spin glass distributions are unstable once one allows for the bare couplings to take generic values $J_{ij} \in \mathbb{R}$. When we allow the disorder PDFs to vary spatially and have them localized to bonds that are sufficiently dilute (defined by requiring sufficient convergence to an IR attractor before disordered bonds start to coalescence), we find a glassy IR attractor corresponding to the Edwards-Anderson $\pm J$ model with equal probability ferro and anti-ferromagnetic bonds.

Upon generalization to two dimensions, single bond disorder is irrelevant. Renormalization group transformations (implemented for tractability through the Migdal-Kadanoff bond shifting technique) rapidly washes out the disorder and restores the bond to the ferromagnetic attractor. The same is true for quenched disorder with a finite number of bonds -- within the sparse limit, disorder is inevitably averaged out. However, for a specific variety of sparse disorder that consists of line defects that wrap, or extend to the system boundary (more generally, codimension one defects), also renormalize like sparse disorder in 1D provided the defects are sparse from the perspective of the orthogonal complement lattice. The relevance of microscopic and mesoscopic disorder in the sparse limit appears to be a feature of codimension one systems alone. In the non-sparse limit, except for special cases corresponding to fixed distributions (which are not attractors in the generic case), probability distribution functions for line defects broaden without limit and the system tends towards a strong disorder regime \cite{Monthus}, which according to the Harris criterion, softens the sharpness of the transition and modifies the critical exponents around the critical point \cite{Harris}.    

As we shall elaborate upon further, the index $\mathcal W$ defined in (\ref{wdef}) directly relates to the Wilson loop operator in higher dimensions (or in one dimension with periodic boundary conditions) \cite{FradkinR}:
\eq{wilson}{W_\gamma = \prod_{i,j \in \gamma} J_{ij}\,;~~~ J_{ij} = \pm 1,}  
where the above is restricted to the special case where the links $J_{ij}$ only take on values $\pm 1$, and where $\gamma$ denotes any closed loop of links. Such models exactly manifest a local $\mathbb Z_2$ invariance that one is tempted to view as a gauge symmetry, with the Wilson loop operator corresponds to a gauge invariant measure of frustration \cite{Toulouse, Fradkin}. The specific assignment $J_{ij} = \pm 1$ however, also allow the Wilson loops to be constructed out of fundamental 'plaquette' variables in two or higher dimensions, 
\eq{plaq}{W_\gamma = \prod_{\square} W_{ijkl}}
where the $W_{ijkl}$ denote a product $\square$ of links around a lattice unit (a plaquette), and where the product is over all plaquettes contained within the curve $\gamma$. This suggests the plaquettes to be the emergent gauge degrees of freedom on the IR attractor. Note that this would not be possible for arbitrary assignments for $J_{ij}$, since the construction (\ref{plaq}) requires the contribution of any disordered bond to adjacent plaquettes to cancel so that the only contributions come from the boundary links, which is only possible if $J_{ij} = \pm 1$. As we show further, for codimension one defects with arbitrarily assigned bond disorder in the bare configuration (i.e. where $J_{ij} \in \mathbb{R}$), which is moreover sparse from the perspective of the orthogonal complement, RG trajectories eventually settle onto an attractor where $J_{ij} = \pm 1$.

The outline of the rest of the paper is as follows -- we begin by considering the renormalization of single bond disorder in 1D and then generalize to multiple bond disorder. We then elaborate on the non-trivial fixed point structure for the disorder parameter and interpret these physically in terms of local twists (or kinks) in the chain that persist at intermediate scales until the RG step where any two coalesce and untwist. We then turn our attention to the 2D case. We find that finite disorder is rapidly washed away under RG transformations, although in the non-sparse limit, qualitatively similar behavior to the 1D case is obtained for line defects due to their being codimension one from the perspective of the complementary lattice. We then comment on obtaining particular cases of the Edwards-Anderson model in the non-sparse limit and discuss how these models represent universality classes for disordered systems that one would flow to for a variety of bare systems. We contextualize our findings in terms of previous studies, comment on the relation to the Kadanoff-Ceva construction and the compatibility of our findings with the Harris criterion for the relevance of disorder to phase transitions before offering our concluding remarks.
	
\section{Sparse disorder in 1D}
\subsection{Preliminaries}
We begin our investigation by considering disorder localized to a single bond in a 1D Ising model and study its behavior under RG transformations. Consider the partition function for a chain with $2N + 1$ sites with free or periodic boundary conditions, indexed $-N \leq i \leq N$ with $N = 2^n$. We leave $N$ finite in the following section, but will consider the $N \to \infty$ limit as needed. In the absence of disorder, the bare partition function is given by
\eq{}{\calZ = \sum_{\{\sigma\}} e^{-\beta_0 H_0}}
where the sum is over all states of the chain, where each site can have spin $\sigma_i = \pm 1$ and where the bare Hamiltonian is given by
\eq{}{-\beta_0 H_0 = \beta_0\sum_{i = -N}^{N-1}\sigma_i\sigma_{i+1} }
i.e., the initial (bare) spin couplings are normalized to unity. Let us now consider a random pair of nearest neighbors, say $m$ and $m+1$ that initially couple to each other with strength $\kappa_0$ instead of unity. The Hamiltonian is now given by
\eq{0ham}{-\beta_0 H_0 = \beta_0\sum_{i \neq m}\sigma_i\sigma_{i+1} + \beta_0 \kappa_0 \sigma_m\sigma_{m+1}}
Defining the bare disorder parameter
\eq{gdef}{g_0 := \kappa_0-1,}
the above can be rewritten as
\eq{h0def}{-\beta_0 H_0 = \beta_0\sum_{i =-N}^{N-1}\sigma_i\sigma_{i+1} + \beta_0 g_0 \sigma_m\sigma_{m+1}}
which is simply the sum of the original Hamiltonian (with identical nearest neighbor interactions) plus an extra contribution parameterizing single bond disorder. 

We would like to consider how this Hamiltonian behaves under renormalization group transformations with each iteration obtained by decimation -- where we fix all even spins (for example) and sum over only the sites for odd $i$. We define
\eq{}{{\rm exp}\left[-\beta_1 H_1(...,\sigma_{-2},\sigma_0,\sigma_2...)\right] = \hspace{-20pt} \sum_{...\sigma_{-3},\sigma_{-1},\sigma_{1},\sigma_3...}\hspace{-20pt} {\rm exp}\left[-\beta_0 H_0\right] }
with $H_0$ given by (\ref{h0def}). Without loss of generality, we presume $m$ to be odd. The summand of the right hand side of the above then factorizes as
\begin{eqnarray}
\label{fact}{\rm exp}\left[-\beta_0 H_0\right]  &=& ...\,e^{\beta_0\sigma_{-1}\left(\sigma_{-2} + \sigma_0 \right)}e^{\beta_0\sigma_{1}\left(\sigma_{0} + \sigma_2 \right)}\\ \nn &\times&...\,e^{\beta_0\sigma_{m}\left(\sigma_{m-1} + [1 + g_0]\sigma_{m+1} \right)}...   
\end{eqnarray}
where we note that all factors are identical except for the one involving the site $m$ and its nearest neighbors. Let's first focus on one of the identical factors not involving $m$, for example:
\eq{}{\sum_{\sigma_1}e^{\beta_0\sigma_1\left(\sigma_0 + \sigma_2\right)} = 2\, {\rm cosh}\beta_0 \left(\sigma_0 + \sigma_2\right) := e^{h(\beta_0) + f(\beta_0)\sigma_0\sigma_2} }
By considering all possible combinations of $\sigma_0 + \sigma_2$, one reproduces
\begin{eqnarray}
\label{notm}
h(\beta_0) &=& \frac{1}{2}{\rm log}\left(4\, {\rm cosh}\, 2\beta_0\right)\\ \nn
f(\beta_0) &=& \frac{1}{2}{\rm log}\left({\rm cosh}\, 2\beta_0\right)
\end{eqnarray}
For site $m$, the last factor in (\ref{fact}) can be written as
\begin{eqnarray}
\nn
\sum_{\sigma_m}&&\hspace{-15pt}e^{\beta_0\sigma_m\left(\sigma_{m-1} + [1 + g_0]\sigma_{m+1}\right)}\\ \nn &=&2\, {\rm cosh}[\beta_0 \left(\sigma_{m-1} + [1 + g_0]\sigma_{m+1}\right)]\\ \nn &:=& e^{\w h(\beta_0,g_0) + \w f(\beta_0,g_0)\sigma_m\sigma_{m+2}}
\end{eqnarray}
Such that similarly, we find
\begin{eqnarray}
\label{m}
\w h(\beta_0,g_0) &=& \frac{1}{2}{\rm log}\left\{4\, {\rm cosh}\, [\beta_0(2 + g_0)]{\rm cosh}\, \beta_0 g_0\right\}\\ \nn
\w f(\beta_0,g_0) &=& \frac{1}{2}{\rm log}\left\{ \frac{{\rm cosh}\, [\beta_0(2 + g_0)]}{{\rm cosh}\, \beta_0 g_0}   \right\}
\end{eqnarray}
writing 
\eq{}{-\beta_1 H_1 = \sum_{i = -2^{n-1}}^{2^{n-1} -1} \left[\beta_1\sigma_{2i}\sigma_{2(i+1)} + c_1 \right] + \beta_1 g_1 \sigma_{m-1}\sigma_{m+1} + d_1, } 
where we've allowed for constant terms localized to the lattice site we've integrated out to be generated by the renormalization group -- in general, all terms not forbidden by the symmetries of the system will be generated by RG flow. We see therefore, that
\begin{eqnarray}
\label{it1}
\beta_1 &=& f(\beta_0)\\
\nn c_1 &=& h(\beta_0)\\ 
\nn g_1 &=& \frac{\w f(\beta_0,g_0) - f(\beta_0)}{\beta_1} = \frac{\w f(\beta_0,g_0)}{f(\beta_0)} - 1\\
\nn d_1 &=& \w h(\beta_0,g_0) - h(\beta_0).
\end{eqnarray}
Upon iteration, one finds
\begin{eqnarray}
\label{it2}
\beta_2 &=& f(\beta_1) = f(f(\beta_0))\\
\nn c_2 &=& h(\beta_1) + 2 c_1 = h(f(\beta_0)) + 2 h(\beta_0)\\ 
\nn g_2 &=& \frac{\w f(\beta_1,g_1) - f(\beta_1)}{\beta_2} = \frac{\w f(\beta_1,g_1)}{f(\beta_1)} - 1\\
\nn d_2 &=& \w h(\beta_1,g_1) - h(\beta_1) + d_1,
\end{eqnarray}
and so forth. The factor of 2 in the second equation arises from taking the sum over half the lattice sites each iteration. We could have also arrived at the same results through the transfer matrix approach \cite{Grinstein}, which we utilize in the appendix in order to calculate spin-spin correlation functions across multiple disordered bonds. We now turn our attention towards the fixed point structure of the renormalization group flow induced by these transformations. 

\subsection{Fixed points and RG trajectories}
\subsubsection{1D Ising model with no disorder}
Consider first Ising 1D chain without any disorder. This is simply implemented in the above by setting the disorder parameter to be vanishing, $g_0 = 0$, implying $\w f = f$ and $\w h = h$. Consequently, we infer
\eq{}{\beta_n = f^n(\beta_0),~~~ f(\beta_0) = \frac{1}{2}{\rm log}\left({\rm cosh}\, 2\beta_0\right)}
Plotting $f(\beta)$ in Fig. \ref{1Db}, we recover the well known result that the only two temperature fixed points in 1D are the trivial stable/unstable fixed points at infinite/zero temperature. We recall here that in the context of the present discussion, our definition of inverse temperature $\beta$ has implicitly absorbed the bare nearest neighbor coupling which we normalized to unity. That is $\beta_0 := J_0/kT_0$ where $J_0$ is the bare nearest neighbor coupling. Hence in 1D, the fixed point at infinite temperature implies the vanishing nearest neighbor interactions in the IR -- were we to plot the renormalized coupling $J = \beta(1 + g)$ in Fig. \ref{phase}, the attractors would merge at $\beta = 0$.

\subsubsection{Ising model with single bond disorder}  
\begin{center}
\begin{figure}[t]\epsfig{file=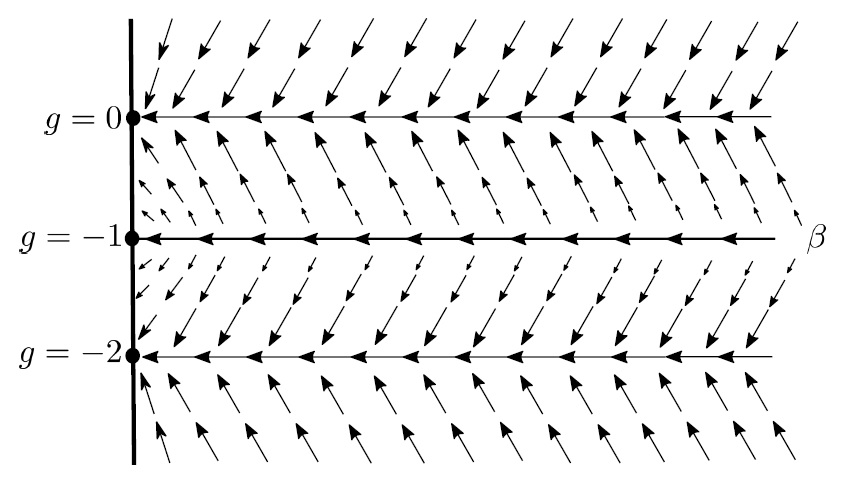, height = 1.9in, width = 3.3in}\caption{\label{phase} Schematic RG flow for single bond disorder.}
\end{figure}
\end{center}

\begin{figure}[t]
	\hfill
	\begin{center}
		\begin{minipage}[t]{.45\textwidth}
			\epsfig{file=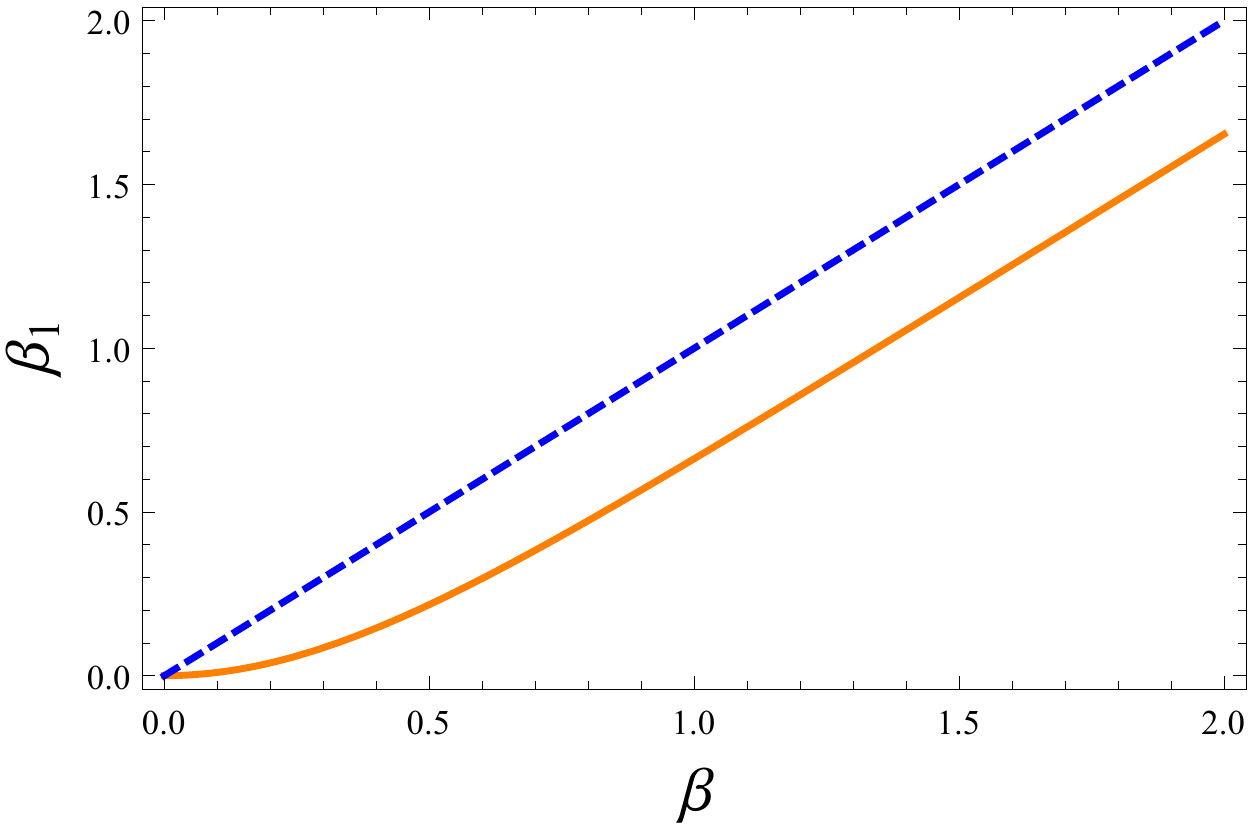, height=2.2in, width=3.1in}
			\caption{\label{1Db} $\beta_1 := f(\beta)$ for the Ising model in the absence of bond disorder.}
		\end{minipage}
		\hfill
		\begin{minipage}[t]{.45\textwidth}
			\epsfig{file=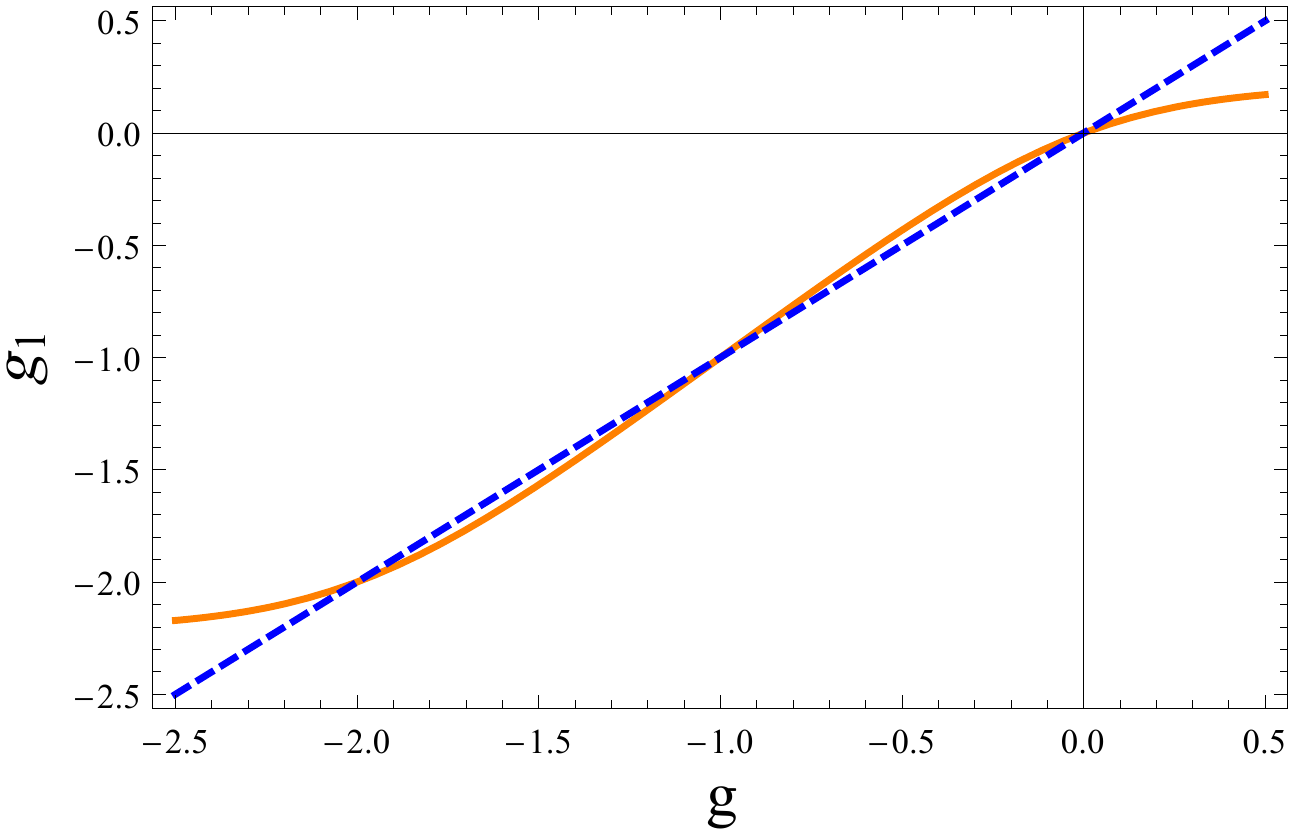, height=2.2in, width=3.1in}
			\caption{\label{1Dg} The function $g_1 := \w f/f -1$ for $\beta = 2$ for the Ising model with single bond disorder.}
		\end{minipage}
	\end{center}
	\hfill
\end{figure}

From the first line of (\ref{it1}) we see that independent of the disorder parameter $g_0$, the behavior of $\beta$ is identical to the ordered case as depicted in Fig. \ref{1Db}. However, the disorder parameter itself has more interesting behavior under RG flow, with fixed points determined by the solutions to the equation
\eq{fp}{g =  \frac{{\rm log}\left\{ \frac{{\rm cosh}\, [\beta(2 + g)]}{{\rm cosh}\, \beta g}   \right\}}{{\rm log}\left({\rm cosh}\, 2\beta\right)} - 1 } 
for some fixed $\beta$. As can be seen by inspection from Fig. 2, and directly from (\ref{fp}) fixed points for the disorder parameter exist for any finite value of $\beta$ at 
\eq{gfp}{g = \{-2,-1,0\},}
corresponding to attractors at $g = -2, 0$ and a separatrix at $g = -1$. Recalling (\ref{gdef}), we see that these possibilities correspond to the nearest neighbor couplings
\eq{}{\kappa = \{-1,0,1\}}
with stable fixed points at $\kappa = \pm 1$ and an unstable fixed point at $\kappa = 0$ ($g = -1$). The former corresponds to the critical line depicted in Fig. \ref{phase} and is equivalent to inserting a break in the chain that is preserved under RG transformations given the nearest neighbor nature of interactions. We note that this feature is unique to 1D – in higher dimensions, next to nearest neighbor and higher order interactions are unavoidably generated. 

We also note that in the infinite temperature limit $\beta \to 0$, the right hand side of (\ref{fp}) becomes $g + \mathcal O (\beta^2)$, so that one is allowed to approach this fixed point with any finite value for the disorder parameter since the effective nearest neighbor interaction given by $\beta(1+g)$ vanishes in this limit.  

Away from the critical line, RG flow directs the disorder parameter to either a stable F or AF attractor depending on the sign of the initial (bare) coupling, corresponding to the fixed points of (\ref{fp}) at $g = \{-2,0\}$ for arbitrary finite temperatures. Away from the critical line at $g = -1$, RG trajectories flow towards the fixed point at $\beta = 0$ along the attractor at $g = 0$ or $g = -2$. 

The fixed point of (\ref{fp}) at $g = -2$ at any finite temperature corresponds to inserting a twist in the original 1D spin chain at the location of the disordered bond in the case of free boundary conditions. This is because physically, given that the interactions are strictly nearest neighbor, flipping the sign of the coupling at a random bond and simultaneously flipping the signs of all spins after that site (say to the left of it) describes an energetically equivalent configuration. Hence the ground state in the presence of single bond disorder corresponds to a kink for any negative bare coupling, no matter its initial magnitude. In the case of periodic boundary conditions, flipping the spins at the disordered bond would eventually result in a frustrated bond somewhere else along the lattice, and the ground state of the spin chain has a degeneracy equal to the number of lattice sites, with all but one ground state corresponding to a double kink. 

As we shall elaborate upon later, random bond models with $\mathbb Z_2$ symmetry exhibit frustration only if the bond configurations are `gauge non-trivial', characterized by the existence of a closed curve with a corresponding Wilson loop (\ref{wilson}) acquiring a non-trivial expectation value \cite{Fradkin,FradkinR}.
\subsection{Arbitrarily assigned bond disorder}
Consider now two disordered bonds, initially separated from each other by some arbitrary distance -- the situation for multiple disordered bonds generalizes straightforwardly. As  before, we write the bare Hamiltonian as
\eq{0hamr}{-\beta_0 H_0 = \beta_0\sum_{i \neq m, l}\sigma_i\sigma_{i+1} + \beta_0 \kappa_0 \sigma_m\sigma_{m+1} + \beta_0 \bar \kappa_0 \sigma_l\sigma_{l+1}.}
Defining individual disorder parameters
\eq{gdefr}{g_0 := \kappa_0-1;~~\bar g_0 := \bar\kappa_0-1}
we see that the above can be rewritten as
\eq{h0defr}{-\beta_0 H_0 = \beta_0\sum_{i =-N}^{N-1}\sigma_i\sigma_{i+1} + \beta_0 g_0 \sigma_m\sigma_{m+1} + \beta_0 \bar g_0 \sigma_l\sigma_{l+1}}
Repeating the previous steps, we find that the decimated Hamiltonian can be written as

\begin{eqnarray}
-\beta_1 H_1 &=& \sum_{i = -2^{n-1} \neq m, l}^{2^{n-1} -1} \left[\beta_1\sigma_{2i}\sigma_{2(i+1)} + c_1 \right]\\ &+& \beta_1 g_1 \sigma_{m-1}\sigma_{m+1} + d_1 + \beta_1 \bar g_1 \sigma_{l-1}\sigma_{l+1} + \bar d_1 \nn
\end{eqnarray}
where
\begin{eqnarray}
\label{it3}
\beta_1 &=& f(\beta_0)\\
\nn c_1 &=& h(\beta_0)\\ 
\nn g_1 &=& \frac{\w f(\beta_0,g_0) - f(\beta_0)}{\beta_1} = \frac{\w f(\beta_0,g_0)}{f(\beta_0)} - 1\\
\nn d_1 &=& \w h(\beta_0,g_0) - h(\beta_0)\\
\nn \bar g_1 &=& \frac{\w f(\beta_0,\bar g_0) - f(\beta_0)}{\beta_1} = \frac{\w f(\beta_0,\bar g_0)}{f(\beta_0)} - 1\\
\nn \bar d_1 &=& \w h(\beta_0,\bar g_0) - h(\beta_0)
\end{eqnarray}
Individually, $g$ and $\bar g$ (in general, any number of disorder parameters) renormalize independently and in an identical functional manner, albeit with different initial conditions, until the RG step where any two nodes coalesce under decimation. At this step, qualitatively different behavior emerges.

\subsection{Bond coalescence under RG}
\begin{figure}[t]\epsfig{file=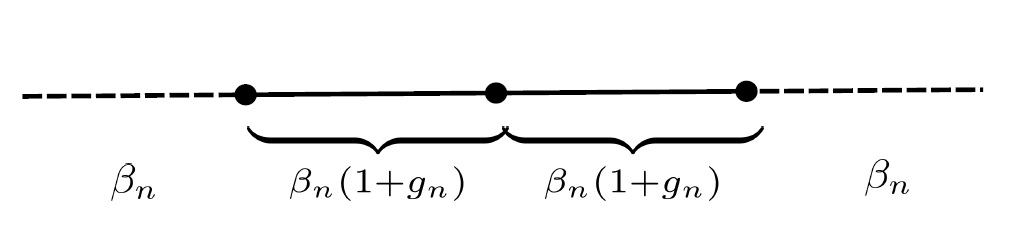, height = 0.9in, width = 3.5in}\caption{\label{RGC} RG step where two nodes coalesce in 1D.}\end{figure} 
Individual disordered bonds evolve independently under RG transformations until the step when they occupy adjacent sites, as illustrated in Fig. \ref{RGC}. The RG transformation that sums over the site connecting two disordered bonds can straightforwardly be shown to result in the renormalized disorder parameter $g_T$, such that
\eq{gT}{g_T(g,\bar g) =  \frac{{\rm log}\left\{ \frac{{\rm cosh}\, [\beta(2 + g + \bar g)]}{{\rm cosh}\,[ \beta (g - \bar g) ]}   \right\}}{{\rm log}\left({\rm cosh}\, 2\beta\right)} - 1.}
Interestingly, for any finite $\beta$, we see that if $g$ or $\bar g$ had settled onto values corresponding to either the F or AF attractor by the step at which the bonds coalesce $g, \bar g = \{0,-1,-2\}$ (\ref{gfp}), the disordered parameter for the single bond $g_{T}(g,\bar g)$ that results is given by
\begin{eqnarray}
&&g_T(0,0) = 0 \label{coalesce1D} \\
&&g_T(0,-1) = -1 \nn\\
&&g_T(0,-2) = -2 \nn \\
&&g_T(-1,-1) = -1 \nn\\
&&g_T(-1,-2) = -1 \nn\\
&&g_T(-2,-2) = 0 \nn	
\end{eqnarray}
We see that anytime either of the adjacent disordered bonds lies on the critical line (i.e. $g, \bar g = -1$) immediately before the RG step where they coalesce, the renormalized $g_T$ is also given by  $g_T = -1$. This is understood from the fact that $g, \bar g = -1$ implies that one of the incoming bonds is of vanishing strength, corresponding to a break in the chain, which is preserved under RG transformations given the nearest neighbor nature of the interactions.

We also note that a twisted bond coalescing with an untwisted bond remains twisted, since $g_T(0,-2) = g_T(-2,0) = -2$, whereas two twisted bonds `untwist' upon coalescence, since  $g_T(-2,-2) = 0$. After coalescence, they renormalized disorder parameter flows according to the single bond RG evolution described by (\ref{fp}) until it coalesces with the next nearest disordered bond. 

The physical picture that emerges is clear -- any bond disorder present in the bare system, no matter how feeble, rapidly settles onto the F or AF attractor depending on its initial sign. Any bonds at the unstable fixed point (or, null bonds) separate the system into independent domains and eliminates the effects of boundary conditions. Disorder persists at intermediate scales until any two twisted bonds meet each other and untwist. At large distances in systems where there are no null bonds, a single defect persists if there are an odd number of initially negative disordered bonds in the bare system, corresponding to the non-vanishing of the index $(-1)^{N_0}$, where $N_0$ is the number of bare negative couplings, resulting in a topological frustrated configuration at long wavelengths in the case of periodic boundary conditions. 

As we shall see in the next section, although sparse disorder is rapidly averaged out by RG transformations in higher dimensions, in the limit of non-sparse disorder the conclusions we have drawn in 1D also apply to codimension one defects that are sparse from the perspective of the complementary lattice. 

\section{Sparse disorder in 2D}
\begin{figure}[t]\epsfig{file=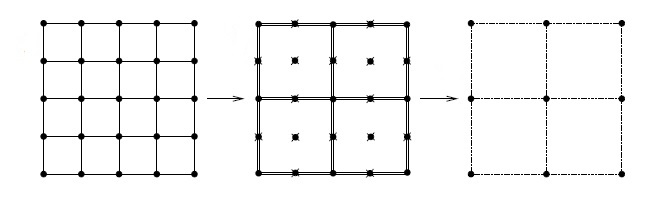, height = 1in, width = 3.3in}\caption{Migdal-Kadanoff bond shifting prescription -- sites marked with an `x' are traced over.\label{mkorg}}\end{figure} 
\begin{figure}[t]
	\hfill
	\begin{minipage}[t]{0.5\textwidth}
	\begin{center}
			\epsfig{file=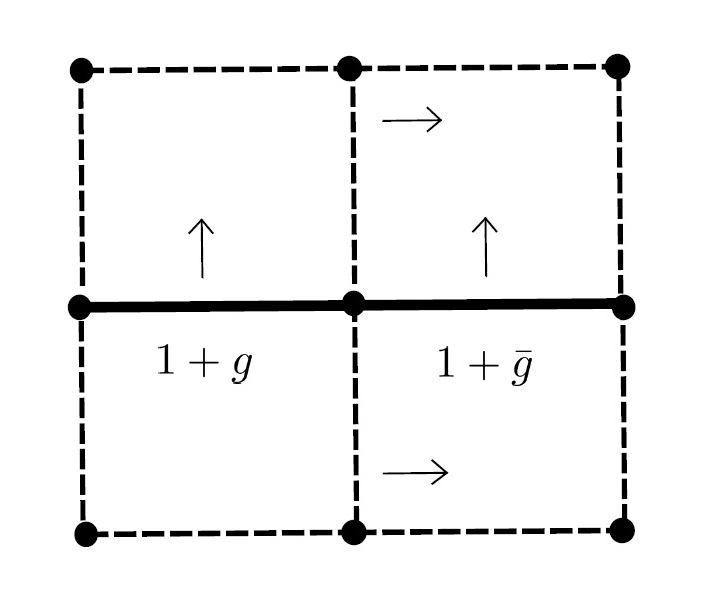, height=1.7in, width=1.9in}
			\caption{\label{shift_sb} MK bond shifting with two disordered bonds.}			
		\end{center}
	\end{minipage}
	\hfill
\end{figure}

\subsection{Migdal-Kadanoff bond shifting}
When considering the renormalization of sparse disorder in two and higher dimensions, one is immediately confronted with the fact that it is significantly harder to implement the renormalization group exactly, forcing us to resort to approximation schemes. For example, the Migdal-Kadanoff (MK) bond shifting scheme and its variants return reasonably accurate answers in comparison to the exact solution in 2D in the case where disorder is absent \cite{Onsager}. The advantage of this approach is that it preserves the nearest neighbor nature of interactions with each RG step and is readily adapted to situations where disorder is present. As illustrated in Fig. \ref{mkorg} one simply shifts every other row of horizontal bonds upwards and every other column of vertical bonds rightwards, decimating over all the mid-point sites. Sites that are left unconnected to any nearest neighbors renormalize the free energy once summed over. 

An improved bond shifting prescription that first decimates and then bond shifts the resulting next to nearest neighbor and four point interactions results in more accurate estimation of the 2D finite temperature fixed point in the absence of disorder. However, even in the presence of single bond disorder, this prescription rapidly proliferates the disorder across multiple bonds quickly rendering the iteration of the RG transformations intractable. Mindful of the many caveats to utilizing the simplest version of MK bond shifting in 2D, in restricting ourselves to studying the scaling behavior and fixed point structure of sparse disorder and disorder that is effectively one dimensional in nature (and sparse from the perspective of the orthogonal compliment), this approximation will turn out to suffice.

\subsection{Renormalization}
We begin by presuming a pair of disordered bonds represented in Fig. \ref{shift_sb}, where after bond shifting and decimating, one finds the renormalized bond disorder parameter
\eq{fp2}{g_1 =  \frac{{\rm log}\left\{ \frac{{\rm cosh}\, [\beta(4 +  g_0 + \bar g_0)]}{{\rm cosh}\, \beta (g_0 - \bar g_0)}   \right\}}{{\rm log}\left({\rm cosh}\, 4\beta\right)} - 1,}
where analogous to Fig. 3, we've parametrized the bond strengths before the RG step as $g_0, \bar g_0$, and denoted $g_1$ the corresponding renormalized value. Decimating over the sites which no longer connect any sites on the renormalized lattice contribute to the renormalized free energy. Proceeding similarly, we find the renormalized temperature fixed point to be determined by the equation
\eq{tfp}{\beta = \frac{1}{2}{\rm log}\left[{\rm cosh} 4\beta \right] }
implying a critical fixed point at $\beta* \approx 0.3$, which is to be compared to the exact value $\beta_0* \approx 0.44$ \cite{Onsager}.
\begin{center}
	\begin{figure}[t]\epsfig{file=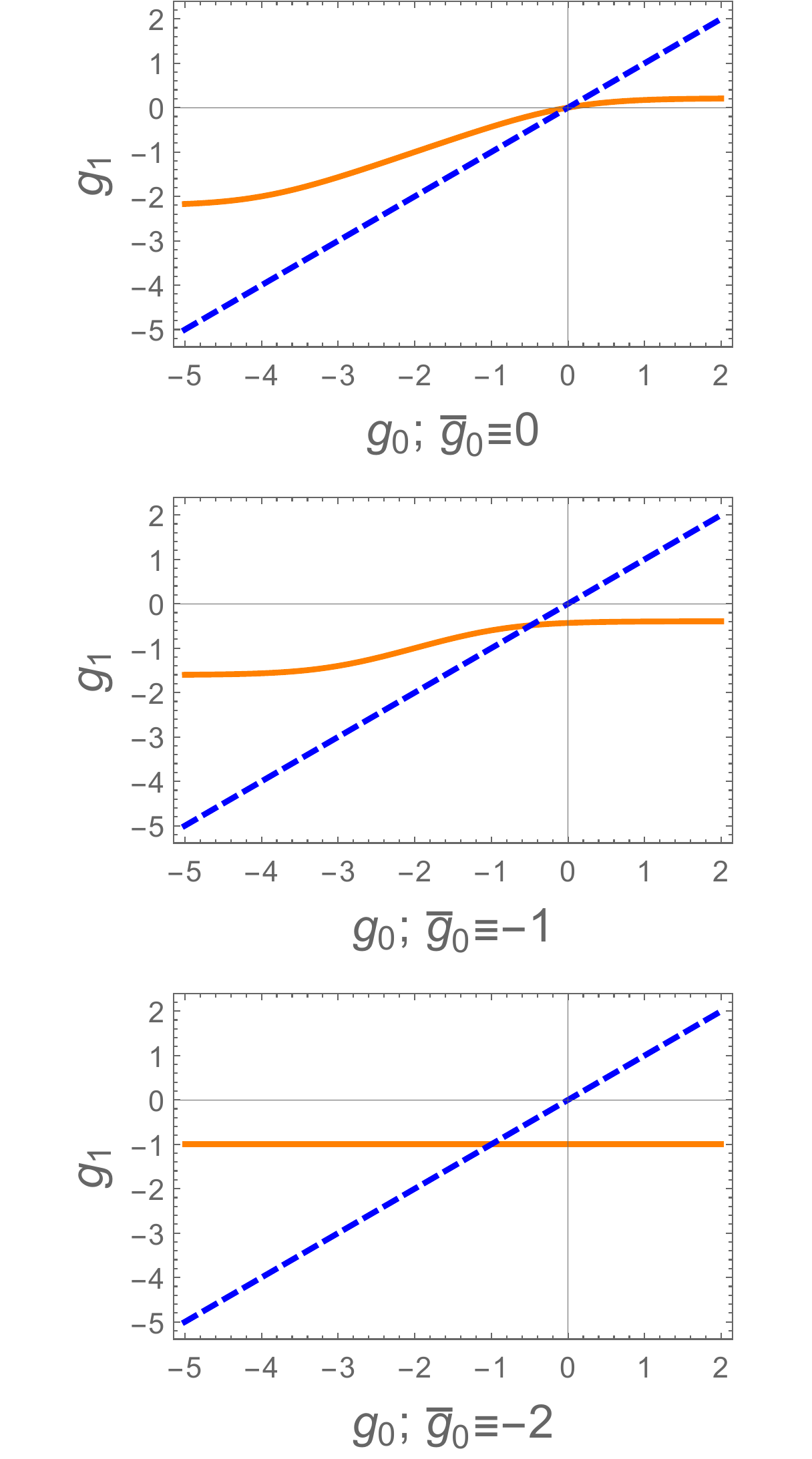, height = 4.7in, width = 2.5in}\caption{Renormalized bond disorder parameter in 2D. \label{2DMK}}
	\end{figure} 
\end{center}
\begin{center}
	\begin{figure}[t]
		\epsfig{file=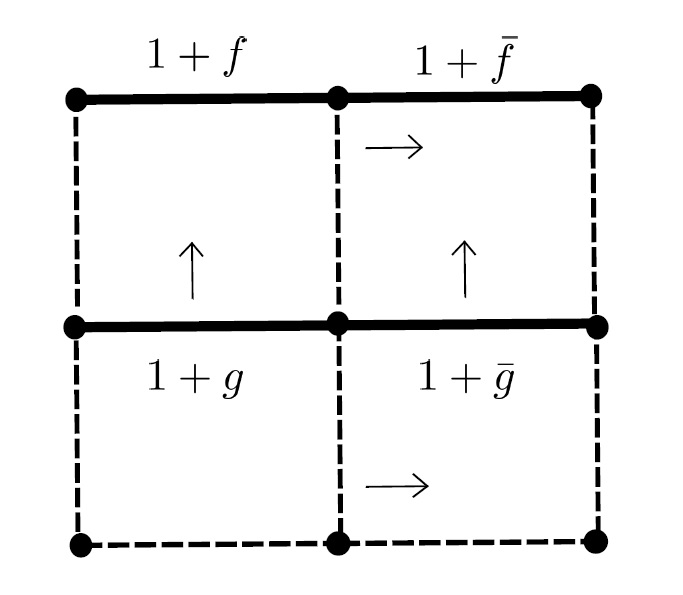, height=1.7in, width=1.9in}
		\caption{\label{shift_db} MK bond shifting with a chain of defects.}
	\end{figure}
\end{center}
\vspace{-50pt}
We see that with a single disordered bond present before any particular RG step (e.g. seen by setting $\bar g_0 = 0$), the tendency is to reduce bond disorder back to $g \to 0$. When one has a second disordered bond which is `twisted' ($\bar g_0 = -2$, so that $\kappa = 1 + \bar g_0 = -1$), the renormalized bond always has vanishing strength. From Fig. \ref{2DMK} we see that successive RG transformations always appear to erase sparse disorder no matter the initial configuration. 

Generalizing the discussion to allow for multiple bond disorder and anticipating the non-sparse limit, we imagine a pair of disordered bonds shifting onto an existing disordered pair (as in fig. \ref{shift_db}). In this case, we find the renormalized bond disorder parameter after decimation to result in
\eq{fp3}{g_1 =  \frac{{\rm log}\left\{ \frac{{\rm cosh}\, [\beta(4 +  g_0 + f_0 + \bar g_0 + \bar f_0)]}{{\rm cosh}\, \beta (g_0 + f_0 - \bar g_0 - \bar f_0)}   \right\}}{{\rm log}\left({\rm cosh}\, 4\beta\right)} - 1 }
One again can similarly deduce from this that for sparse bond disorder, subsequent RG transformations rapidly renders localized disorder irrelevant. This needn't be the case for infinite numbers of defects, whether or not they have a vanishing density in the thermodynamic limit. 

\section{Spatially extended sparse and non-sparse disorder}

\subsection{Line defects}

\begin{figure}[t]
	\epsfig{file=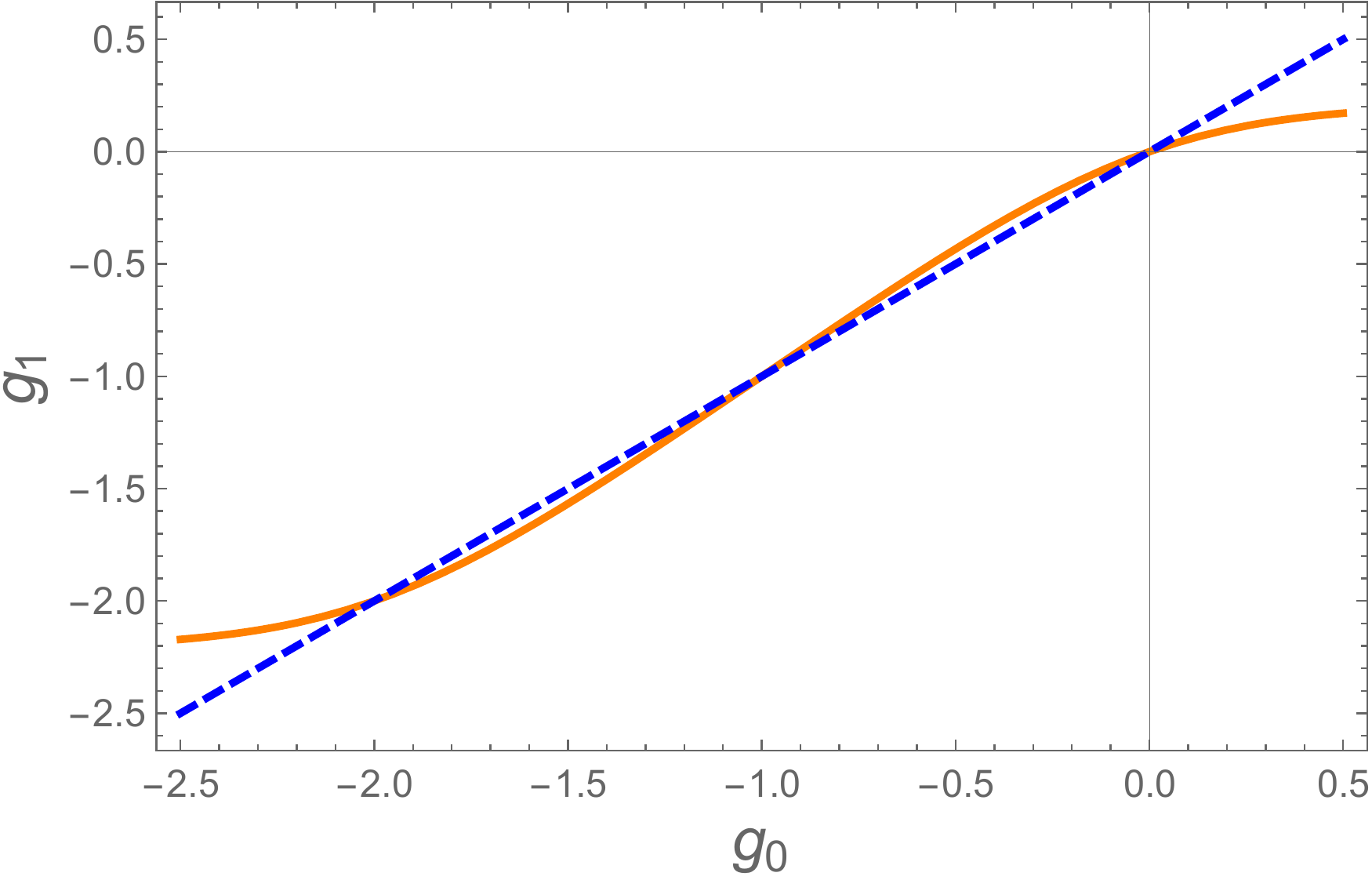, height=2.2in, width=3.1in}
	\caption{\label{2dld} The renormalization of the disorder parameter for a line defect in 2D.}
\end{figure}

One can also study the case where disorder is no longer localized to small regions. Consider a 2D lattice with a line defect, which we model as a vertical column of horizontally oriented disordered bonds with identical disorder parameter. When periodic boundary conditions are imposed, this line is taken to form a complete cycle along one direction. One can immediately infer how this defect is renormalized by considering (\ref{fp3}), setting $g_0 = f_0 $ and $\bar g_0 = \bar f_0 = 0$, and considering the continued iteration of this formula
\eq{2dfp}{g_1 =  \frac{{\rm log}\left\{ \frac{{\rm cosh}\, [\beta(4 +  2 g_0)]}{{\rm cosh}\, 2\beta g_0}   \right\}}{{\rm log}\left({\rm cosh}\, 4\beta\right)} - 1.}
Unsurprisingly, since the geometry of the disorder is translationally equivalent  to a single disordered bond in 1D, this has the same fixed point structure as the 1D case, as can be seen from Fig. \ref{2dld} with the added feature that in 2D one also has a non-trivial unstable temperature fixed point at $\beta_* = 0.44$. Since the fixed points of (\ref{2dfp}) at $g = \{-2, -1, 0\}$ occur for any finite value of $\beta$, the disorder parameter flows towards its nearest attractor regardless of which side of the  critical point one is, although this convergence is much more rapid at lower temperature ($\beta > \beta_*$). As one approaches $\beta \to \infty$, the presence of relevant disorder at long wavelengths will be macroscopically distinguishable.

We can also examine the renormalization of a discrete collection of line defects. As in the 1D case for sparse bond disorder, by locality of interactions, individual line defects renormalize independently until the RG step where they coalesce. At  this stage, the renormalized line defect disorder parameter gets renormalized as 
\eq{ldren}{g_T =  \frac{{\rm log}\left\{ \frac{{\rm cosh}\, [\beta(4 +  2g_0 + 2\bar g_0)]}{{\rm cosh}\, 2\beta (g_0 - \bar g_0)}   \right\}}{{\rm log}\left({\rm cosh}\, 4\beta\right)} - 1 }
obtained from (\ref{fp3}) by setting $g_0 = f_0$ and $\bar g_0 = \bar f_0$. We see that analogous to the 1D case (\ref{coalesce1D}), upon defect coalescence, the renormalized defect disorder parameter is given by
\begin{eqnarray}
&&g_T(0,0) = 0 \label{coalesce2D} \\
&&g_T(0,-1) = -1 \nn\\
&&g_T(0,-2) = -2 \nn \\
&&g_T(-1,-1) = -1 \nn\\
&&g_T(-1,-2) = -1 \nn\\
&&g_T(-2,-2) = 0. \nn	
\end{eqnarray}
This again implies that any two line defects that have settled onto the AF attractor untwist from the perspective of the orthogonal complement lattice at the RG step where they coalesce. When one considers more complicated configurations of line defects that may intersect in 2D, one would find that the individual defects renormalize independently. Moreover, one does not add any local frustration with such configurations since it is not possible to construct a closed loop that encloses the intersection without traversing an even number of disordered bonds.

\subsection{Random bond disorder and the Edwards-Anderson limit}

The Edwards-Anderson model is defined by the bare Hamiltonian \cite{EA} 
\eq{EAham}{H = \sum_{\langle i j \rangle} J_{ij}\sigma_i\sigma_j}
where the sum is over nearest neighbor lattice sites. The weights $J_{ij}$ are taken to be drawn from some PDF, denoted $P(J_{ij})$. For the Gaussian Edwards-Anderson model, one has 
\eq{}{P(J_{ij}) = \frac{1}{\sqrt{2\pi J^2}}\,{\rm exp}\left[-\frac{(J_{ij} - J)^2}{2J^2}\right]}
with mean $J_0$ and variance $J^2$. The so called $\pm J$ Edwards-Anderson model is defined by the PDF
\eq{jea}{P(J_{ij}) = p\,\delta(J_{ij} - J) + (1-p)\,\delta(J_{ij} + J)}
where the bonds have bare strength $J$ or $-J$ with probabilities $p$ or $1-p$ respectively. 

The renormalization of the Edwards-Anderson model can be obtained by considering the non-sparse limit of our previous results (\ref{gT}) and (\ref{fp3}) by assigning to each bond, a bare disorder parameter drawn from the corresponding PDF. The main difference with the cases studied in previous sections is that each step in the RG involves node coalescence. Via the replica trick, one could compute the free energy, various order parameters and the ground states of the system. We note however that the standard replica trick ceases to apply to sparse disorder however, since one is in effect assigning different PDF's for the bond disorder at specific individual bonds. However, one can still straightforwardly infer how the initial PDF's renormalize in either case, as we shall do shortly.

Considering first non-sparse disorder in 1D, and recalling that each RG step involves node coalescence, two adjacent bonds with disorder parameters $g_1$ and $g_2$ result upon decimation in the renormalized disorder parameter $g_T$ (\ref{gT}):
\eq{gT2}{g_T(g_1,g_2)  =  \frac{{\rm log}\left\{ \frac{{\rm cosh}\, [\beta(2 + g_1 + g_2)]}{{\rm cosh}\,[ \beta (g_1 - g_2) ]}   \right\}}{{\rm log}\left({\rm cosh}\, 2\beta\right)} - 1}
The PDF for the renormalized disorder parameter, denoted $P_1(g)$, given the PDF for the initial disorder $P_0(g)$ is obtained as
\eq{p1}{P_1(g) = \int\,dg_1\,dg_2\,P_0(g_1)P_0(g_2)\delta(g_T(g_1,g_2) - g)} 
It may be tempting to perform one of the integrals via the delta function, however the symmetric function $g_T(g_1,g_2)$ fails to be invertible when either $g_1$ or $g_2 = -1$. We could consider the condition where the PDF's preclude this possibility, however it would not correspond to the most general case. One can nevertheless proceed by studying the integral equation condition implied by demanding the PDF be invariant, that is a \textit{fixed distribution} under RG transformations --
\eq{}{P_*(g) = \int\,dg_1\,dg_2\,P_*(g_1)P_*(g_2)\delta(g_T(g_1,g_2) - g)}   
which is satisfied whenever
\eq{}{\int\,dg'\,P_*(g')\delta(g_T(g_1,g') - g) = \delta(g_1 - g)}
which implies that $P_*(g)$ must itself be a delta function, or a weighted sum thereof. 

We can recover the fixed distributions found by Grinstein et al \cite{Grinstein} by considering an initial distribution of the functional form reasoned above,
\eq{id}{P_0(g) = p_0\,\delta(g) + (1-p_0)\,\delta(g+2),~~ 0 \leq p_0 \leq 1,}
and iterating this under RG. Substituting (\ref{id}) into (\ref{p1}),  performing the integrations over $g_1$ and $g_2$ and using the relations (\ref{coalesce1D}), we find
\begin{eqnarray}
P_1(g) &=& \left(p_0^2 + (1-p_0)^2\right)\,\delta(g) + 2p_0(1-p_0)\,\delta(g+2) \nonumber\\
&\equiv& p_1\,\delta(g) + (1-p_1)\,\delta(g+2)
\end{eqnarray}
with 
\eq{fprob}{p_1 := p_0^2 + (1-p_0)^2.} 
Evidently, the fixed distributions are determined by the fixed points of the equation $p_* = p_*^2 + (1-p_*)^2$, which are at $p_* = 1$ and $p_* = 1/2$. 

The fixed point at $p_* = 1$ implies the absence of disorder in the initial configuration. \textit{However, this fixed point is unstable}, as iterating (\ref{fprob}) for any initial values for $p_0$ slightly less than unity, parametrized as $p_0 = 1 - 2^{-n}$, can straightforwardly be verified to converge on $p_* = 1/2$ in approximately $n+1$ RG steps. This has a transparent physical interpretation -- (\ref{coalesce1D}) implies that any AF bond coalescing with an F bond at a given RG step renormalizes to an AF bond. Therefore AF bonds that are extremely rare initially `eat up' neighboring F bonds under RG transformations until they encounter another AF bond (at which point they renormalize to an F bond). Hence, a statistical steady state can't be reached until the RG step where the average separation between renormalized AF bonds equals that of the renormalized F bonds, which is after approximately $n+1$ decimations if one had initially assigned the probability of finding an AF bond to be $2^{-n}$. 

We find however that once we allow for a finite probability for null bonds within the restricted class of PDF's consisting of sums of delta functions (i.e., allowing $g = -1$ with some finite probability $q$), the fixed point for the distribution (\ref{id}) at $p_* = 1/2$ also becomes unstable. Consider 
\eq{new}{P_0(g) = p_0\,\delta(g) + q_0\,\delta(g+1) + (1-p_0 - q_0)\,\delta(g+2),}
with $p_0, q_0 \leq 1$ and $p_0 + q_0 \leq 1$, we find upon iterating the RG transformation, the distribution $P_1(g)$ given by (\ref{p1}) has the same form as (\ref{new}) but now with probability assignments
\begin{eqnarray}
p_1 &=& p_0^2 + (1 - p_0 - q_0)^2 \\ \nonumber 
q_1 &=& 2q_0(2-q_0)
\end{eqnarray}
which can be shown to rapidly converge to $q_* = 1, p_* = 0$ for any finite value of $q_0$. This `paramagnetic' fixed distribution is trivial, and corresponds to a completely disconnected spin chain and is the only attractor among the fixed distributions found in \cite{Grinstein}. That this should be so is evident from the fact that any break in the chain is preserved under renormalization, and so any bond with disorder parameter $g = -1$ will eat up all adjacent bonds under successive RG steps. 

We stress that the conclusions drawn above are an artifact of the specific form for the bare PDF given by (\ref{new}), wherein the arguments of the delta functions corresponds to fixed points of the disorder parameter (\ref{gT2}). By allowing for an initial distribution of the form 
\eq{kap}{P_0(g) = p_0\,\delta(g - \kappa_1) + (1-p_0)\,\delta(g -\kappa_2),~~ 0 \leq p_0 \leq 1}
with $\kappa_1, \kappa_2 $ taking generic values $\neq \{0,-1,-2\} $, a single iteration of the RG results in a weighted sum of three delta functions with arguments $g_T(\kappa_1,\kappa_1), g_T(\kappa_1,\kappa_2)$ and $g_T(\kappa_2,\kappa_2)$, each different from each other. Repeating this process, one gets a weighted sum of six delta functions with different arguments given by the symmetric combinations of $g_T$ acting on the results from the previous step: each RG transformation generates $N(N+1)$ more terms from the $N$ generated in the previous step. Moreover, we see that setting both arguments to be identical in (\ref{gT2}) for $g_1 = g_2 > 0$ or $< -2$, the width of the distribution gets successively wider. However when one allows $\kappa_1 = -1$ but let $\kappa_2$ to take on any generic value in (\ref{kap}), RG transformations preserve the form of the PDF to be a sum of only two delta functions where the paramagnetic fixed distribution has a non-trivial basin of attraction provided $-2 < \kappa_2 < 0$. 

\begin{figure}[t]
	\epsfig{file=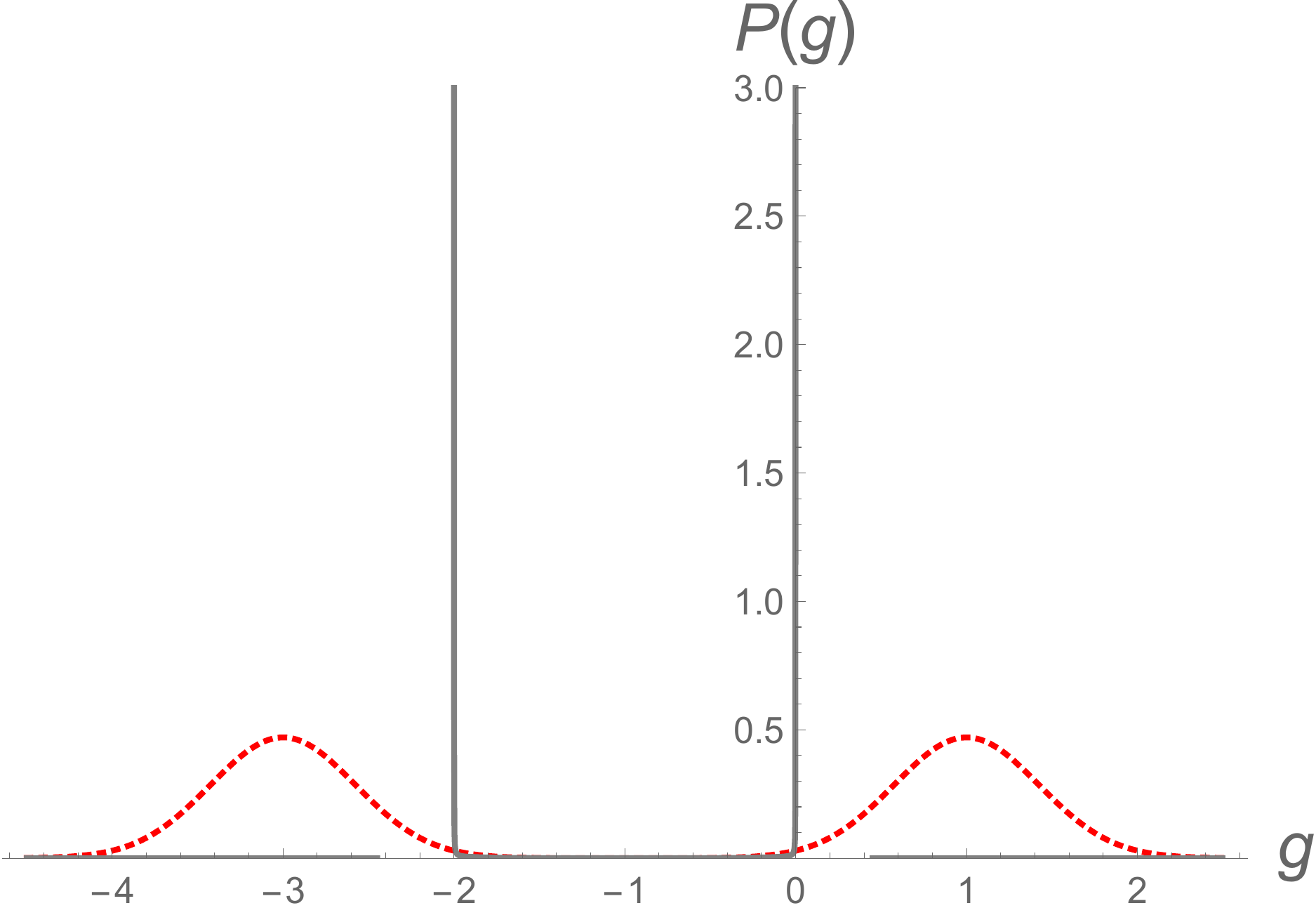, height=2.2in, width=3.1in}
	\caption{\label{pdf0} Bare PDF that is the sum of two distributions of the form (\ref{gpdf}) with $\sigma = 0.6$, centered around $\bar g = 1$ and $\bar g = -3$ (red dashed line) after one RG iteration.}
\end{figure}

\begin{figure}[t]
	\epsfig{file=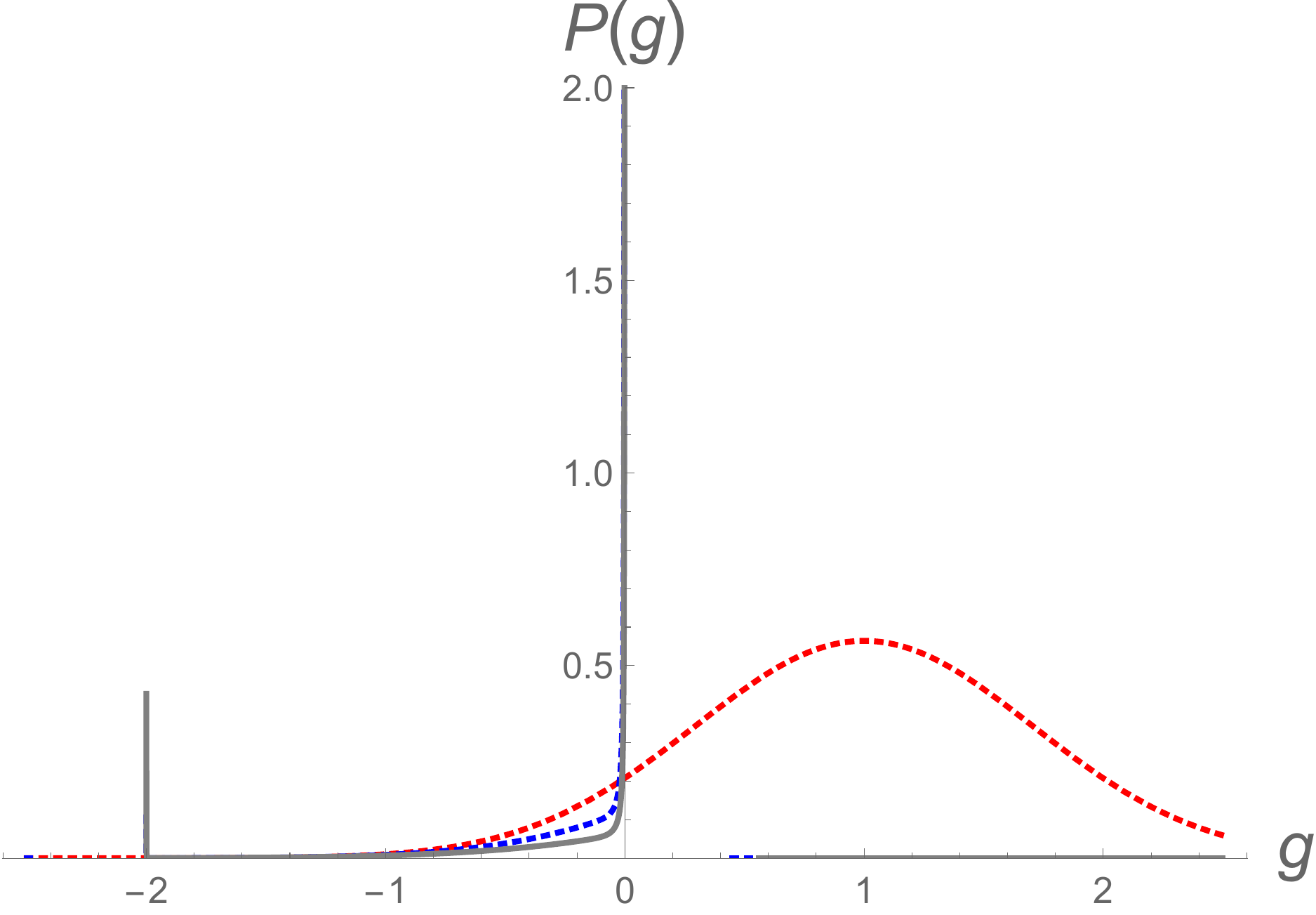, height=2.2in, width=3.1in}
	\caption{\label{pdf1} The bare PDF (\ref{gpdf}) with $\sigma = 1$, $\bar g = 1$ (red dashed line) after 12 (blue) and 24 RG (gray) iterations.}
\end{figure}

\begin{figure}[t]
	\epsfig{file=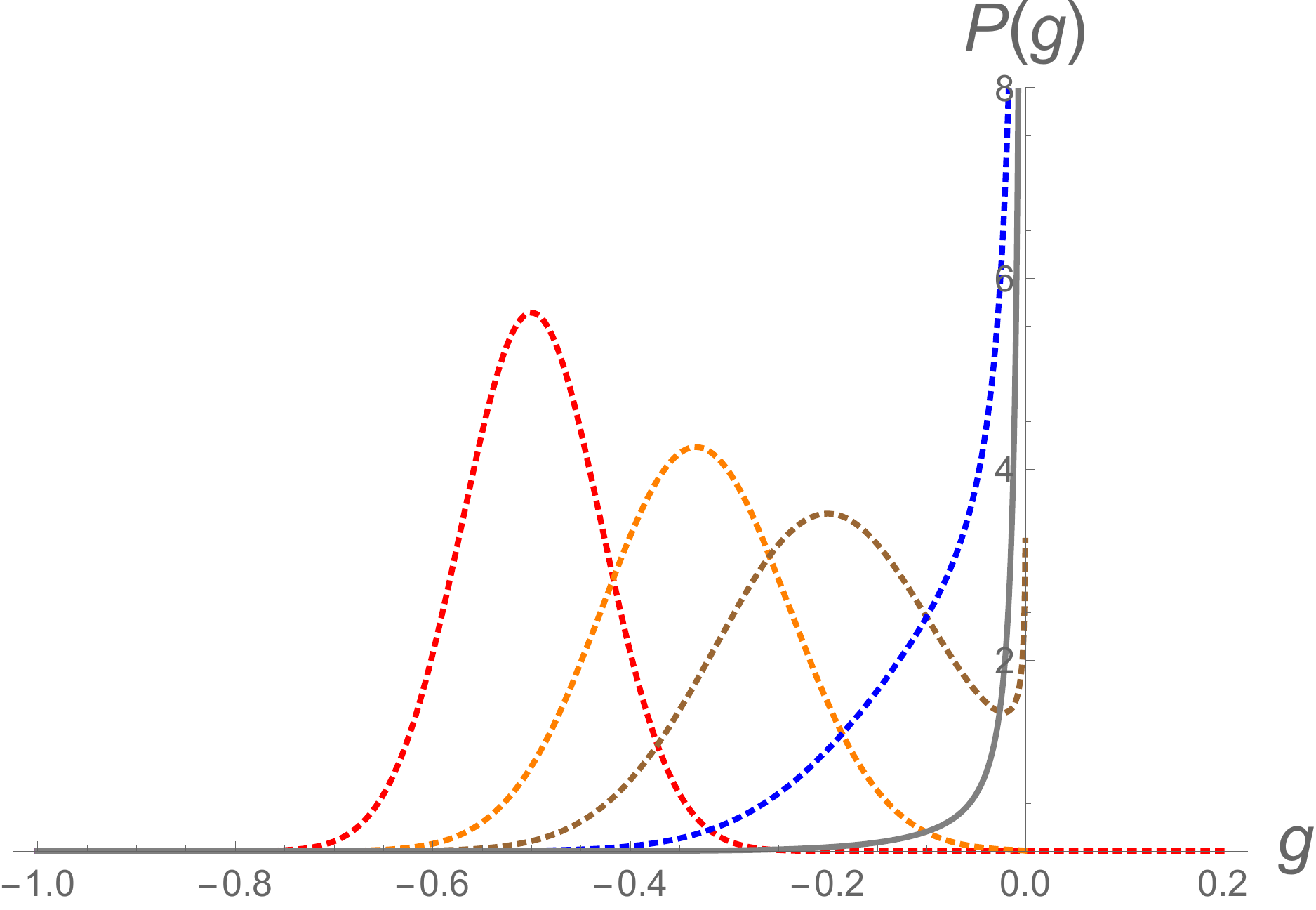, height=2.2in, width=3.1in}
	\caption{\label{pdf2} The bare PDF (\ref{gpdf}) with $\sigma = 0.1$, $\bar g = -0.5$ (red dashed line) after 18, 27, 36 and 45 RG iterations (orange, brown, blue, and gray, respectively).}
\end{figure}

\begin{figure}[t]
	\epsfig{file=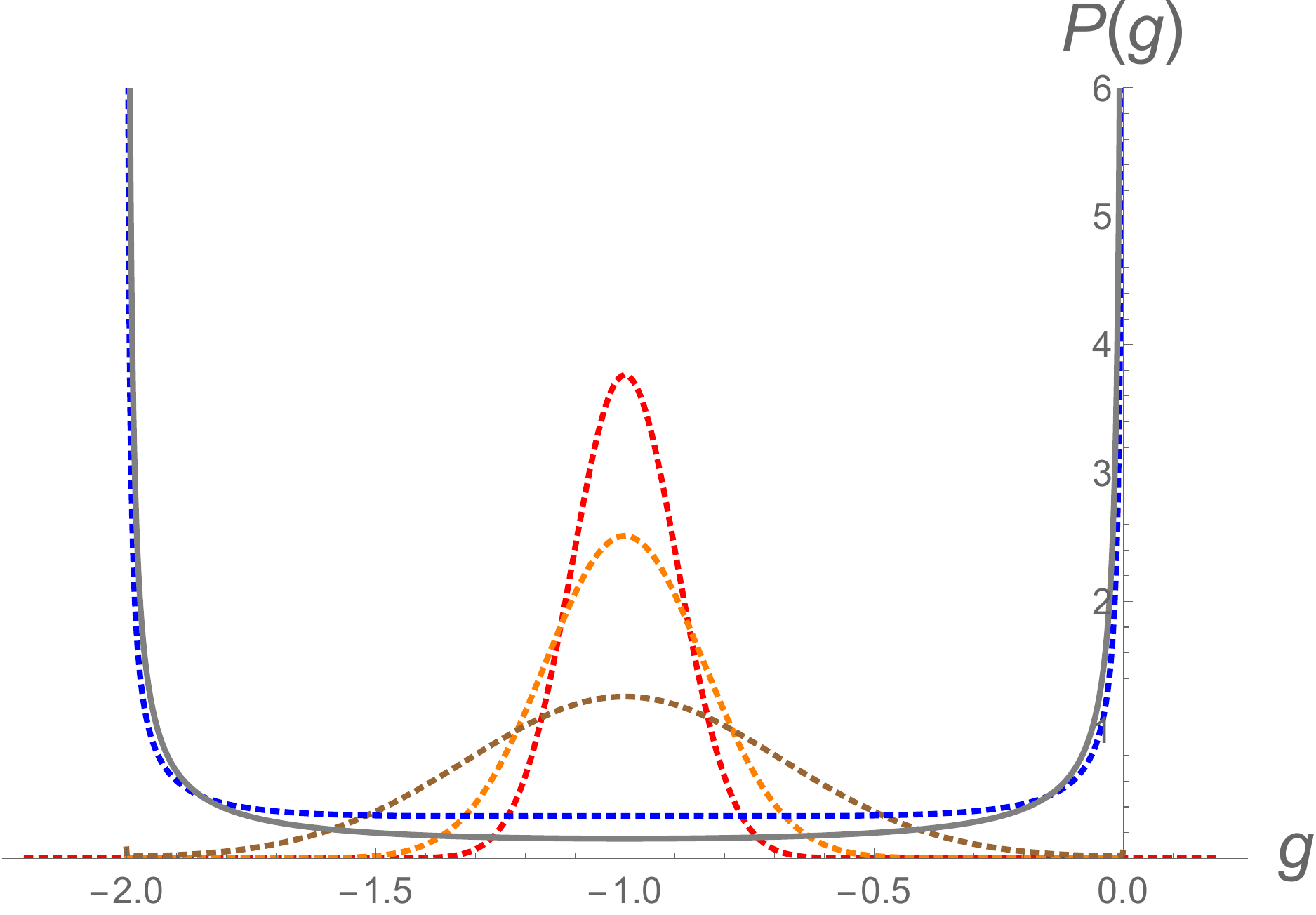, height=2.2in, width=3.1in}
	\caption{\label{pdf3} The bare PDF (\ref{gpdf}) with $\sigma = 0.15$, $\bar g = -1$ (red dashed line) after 24, 48, 66 and 72 RG iterations (orange, brown, blue, and gray, respectively).}
\end{figure}

The situation becomes richer when we allow non-sparse disorder to be inhomogeneous. In this case, allowing for two adjacent disordered bonds to have disorder parameters drawn from \textit{different} PDF's, we find that the renormalized PDF is now given by
 \eq{p3}{P_1(g) = \int\,dg_1\,dg_2\,P_0(g_1)P'_0(g_2)\delta(g_T(g_1,g_2) - g)} 
with $P_0(x) \neq P_0'(x)$. We consider the simple case where the disorder is diluted enough such that there are a finite number of RG steps before coalescence with another disordered bond occurs. In this case, $P_0'(g) = \delta(g)$, so that the renormalized PDF becomes
\eq{pdfren2}{P_1(g) = \int\,dg_1\,P_0(g_1)\delta(g_T(g_1,0) - g)}
The function $g_T(g,0)$ is monotonic and hence invertible. Therefore
\eq{pdfren}{P_1(g) = P_0[\gamma^{-1}(g)]\bigg/\frac{d \gamma}{d g}\bigg|_{\gamma^{-1}(g)}}
with
\eq{gamma}{\gamma(\beta, g) \equiv g_T(0,g) = \frac{{\rm log}\left\{ \frac{{\rm cosh}\, [\beta(2 + g)]}{{\rm cosh}\, \beta g}   \right\}}{{\rm log}\left({\rm cosh}\, 2\beta\right)} - 1,}
which has the inverse
\eq{}{\gamma^{-1}(\beta, g) = \frac{1}{2\beta} \log\left[\frac{(\cosh\,2\beta)^{1 + g} - e^{-2\beta}}{e^{2\beta} - (\cosh\,2\beta)^{1 + g}}\right].}
One can iterate the transformation (\ref{pdfren}) up till the RG step where bond coalescence occurs. Considering a Gaussian bare PDF for the disorder parameter
\eq{gpdf}{P(g) = \frac{1}{\sqrt{\pi \sigma^2}}\,{\rm exp}\left[-\frac{(g - \bar g)^2}{\sigma^2}\right]}
We see from Figs. \ref{pdf0} - \ref{pdf3} that the PDF's tend towards the distribution of the form (\ref{id}). This convergence is slower for bare PDF's with more support around the unstable fixed point at $g = -1$, however is rather rapid when this is not the case, cf. Fig. \ref{pdf0}, which depicts a single iteration of the RG. This can be understood from the stiffness (flatness) of the function $\gamma(\beta,g)$ for larger absolute values of $g$ at any finite temperature (as opposed to near linearity around $g = -1$). The decimation of the spin in between a bond with say $g = 1$ and $g \gg 1$ results in an effective interaction that is still $\mathcal O(1)$, with renormalization primarily shifting the excess energy associated with the highly disordered bond into a local contribution to the free energy. Physically this is because although it is energetically favorable for the intermediate spin to align with the adjacent spin that it is strongly coupled to, the coupling to the spin connected via an ordered bond remains the same, such that the effective coupling across the intermediate spin is still $J = 1+g$ with $g \sim \mathcal O(1)$. In other words, a chain is only as strong as its weakest link.  

Provided the non-sparse disorder has a PDF that converges sufficiently to the distribution (\ref{id}) before bond coalescence onsets (which occurs within a small number of RG iterations if the bare PDF lacks significant support around the unstable fixed point at $g = -1$), subsequent RG transformations involving bond coalescence will renormalize the distribution
\eq{}{\nonumber P(g) = p\,\delta(g) + (1-p)\,\delta(g+2)}    
which converges to the fixed distribution $P_*$ with $p_* = 1/2$. We can therefore conclude that the $\pm J$ Edwards-Anderson model (\ref{jea}) with $p = 1/2$ represents a universality class for disordered systems in 1D, which any model within a non-trivial basin of attraction will flow towards. Note that the '$\pm J$' terminology describes the unnormalized coupling $J = \beta(1 + g)$. 

For line defects in two dimensions, one can straightforwardly repeat the analysis with the replacement
\eq{}{\bar{\gamma}(\beta,g) = \frac{{\rm log}\left\{ \frac{{\rm cosh}\, [\beta(4 +  2 g)]}{{\rm cosh}\, 2\beta g}   \right\}}{{\rm log}\left({\rm cosh}\, 4\beta\right)} - 1}
in place of $\gamma(\beta,g)$ in (\ref{pdfren}) and arrive at qualitatively similar conclusions. The situation is far richer in 2D however due to the existence of a finite temperature critical point. The case of non-sparse, statistically distributed disorder has been extensively studied (see \cite{Monthus} for an excellent review) and the effect of disorder on the critical points of the system depends on whether or not the Harris criterion is satisfied \cite{Harris} -- that is, if the exponent $\nu$ of a correlation length associated with a given critical point satisfies $\nu \geq d/2$. This occurs when the width of the disorder PDF's grows without bound under RG, which, as we found previously, will be the case for homogeneously assigned disorder. For inhomogeneously assigned bare disorder where the Edwards-Anderson attractor is reached sufficiently rapidly, the critical properties of the system are unchanged.

\section{Aspects of $\mathbb Z_2$ gauge symmetry in the IR \label{Z2}}
The Ising model with sparse disorder furnishes a concrete example of a system in which gauge degrees of freedom naturally manifest in the IR. Whether or not gauge degrees of freedom can be emergent in continuum models is a question of great interest to particle physics \cite{IAT, JD, EW, CB}, motivating the study of any condensed matter examples where this occurs \cite{Wen}.       

Recall the Hamiltonian (\ref{EAham}), that is, a random bond model which needn't for the purposes of the present discussion be drawn from a PDF. Evidently, the Hamiltonian admits an invariance under the transformation 
\eq{z2}{\sigma_i \to -\sigma_i;~J_{ij} \to -J_{ij},~~ \forall j {\rm~ adjacent~to~ } i}
for certain fixed configurations of bonds. In this case, one finds the energetics, and therefore the thermodynamic properties of the disordered lattice to be identical to the completely ordered case (\ref{z2}) \cite{Toulouse}. 

Configurations that cannot be related to the ordered Hamiltonian via local transformations of the form (\ref{z2}) have different thermodynamic and ground state properties and exhibit frustration at long wavelengths \cite{Fradkin, FradkinR}. Viewing the transformation (\ref{z2}) as a local $\mathbb Z_2$ invariance, one can show that for any closed loop of bonds, the quantity
\eq{wilson2}{W_\gamma = \prod_{i, j \in \gamma} J_{ij}\,;~~~ J_{ij} = \pm 1}  
is a $\mathbb{Z}_2$ invariant measure of frustration, with $W_\gamma = -1$ implying the existence of at least one frustrated bond. 

The results of the previous sections imply that the ever present $\mathbb Z_2$ invariance (\ref{z2}) manifests in the specific form (\ref{wilson2}) once all disordered bonds have settled onto one of two RG attractors (corresponding to $g = -2, 0$). That is, even if we start with a bare configuration such that 
\eq{reals}{J^{\rm bare}_{ij} \in \mathbb R,} 
renormalization group flow results in long wavelength configurations with $J_{ij} \to \pm 1$ for a range of bare disordered configurations. These correspond to sparse disorder, or non-sparse disorder that is inhomogeneously distributed with sufficiently rapid convergence to the Edwards-Anderson $\pm J$ universality class, which will be the case for any initial distributions with sufficiently vanishing support around the critical line at $g = -1$. 

The index $\mathcal{W} = (-1)^{N_0}$ where $N_0$ counts the number of bonds with bare disorder parameters $g_0 < -1$ is invariant under RG transformations, and corresponds to the Wilson loop (\ref{wilson2}) for a large enough loop (i.e. in the IR). This specific distribution allows one to construct all quantities of thermodynamic interest from fundamental plaquette variables (not possible with the distribution (\ref{reals}) since the plaquette construction requires contributions from disordered bonds common to adjacent plaquettes to cancel), implying the former to be the relevant gauge degrees of freedom in the IR.

Configurations with $W_\gamma = -1$ for some $\gamma$ have disorder that is relevant at the largest scales, corresponding to a topologically frustrated configuration in the case of periodic boundary conditions. In the non-sparse limit, long wavelength configurations with $J_{ij} \to \pm 1$ correspond to a particular realization of the random bond model (\ref{jea}) whose phase structure in 2D was first explored in \cite{CF}. Kadanoff and Ceva \cite{KC} have furthermore shown that if $\tilde r$ and $\tilde r'$ denote two sites on the dual lattice (dual to $r$ and $r'$) that are connected by a path of disordered bonds denoted $\Gamma$, then one can define a \textit{disorder operator} $\mu$ defined via its correlation function
\eq{}{\langle \mu(\tilde r)\mu(\tilde r') \rangle := \frac{\mathcal Z[\Gamma]}{Z} \equiv e^{-\frac{\Delta F[\Gamma]}{T}},} 
where $\mathcal Z[\Gamma]$ is the partition function with an insertion of a seam of flipped bonds $\Gamma$. With the identification of the Kramers-Wannier dual coupling  
\eq{}{e^{-2K^*} = {\rm tanh}\, K;~~~ K := J/T}
the correlation functions of the disorder operator in the dual system can be shown to be equal to the spin correlations in the original model \cite{KC}
\eq{}{\langle\sigma(r)\sigma(r') \rangle_K = \langle\mu(\tilde r)\mu(\tilde r') \rangle_{K^*}} 
and that the Wilson loop $W_\gamma = -1$ only if the one (but not both) of the disorder operators $\mu(r)$ lie within the loop, i.e. the seam $\Gamma$ pierces the loop. As before, frustration is only present when $W_\gamma = -1$.
\\

\section{Concluding remarks}
In this investigation, we have explored how sparse disorder behaves under renormalization. We found that in 1D, arbitrarily assigned bare disorder localized to a finite number of bonds flows towards an attractor where individual bonds settle onto F or AF couplings of equal and opposite strength. In 2D this corresponds to plaquettes emerging as the relevant gauge degrees of freedom in the IR. Disorder is relevant at large scales if the index $(-1)^{N_0}$ associated with the bare configuration (where $N_0$ is the number of bonds with bare disorder parameter $g_0 < -1$) is non-trivial, corresponding to the Wilson loop operator $W_\gamma$ in the IR. 

In the limit of non-sparse disorder that is homogeneous, we recover known fixed distributions, but find only the trivial paramagnetic fixed distribution to be an attractor when allowing for the disorder parameter to take on arbitrary values not just restricted to $g = \{-2,-1,0\}$ i.e. the fixed points of (\ref{fp}). On the other hand, inhomogeneous non-sparse disorder with sufficiently vanishing support for the disorder parameter around the critical line $g = -1$ flows towards the Edwards-Anderson equal probability $\pm J$ model in the IR.

Localized sparse disorder in 2D is rapidly washed out under renormalization. However, line defects that are sparse from the perspective of the orthogonal complement lattice behave analogously to sparse disorder in 1D, implying the generalization of our results to codimension one defects in 2 and higher dimensions. 

\section{Acknowledgements}

We wish to thank Poul Henrik Damgaard, John Donoghue, Mehran Kardar, David Sherrington, Robin Stinchcombe and especially Mike Schecter for valuable discussions and comments. SP is supported by funds from Danmarks Grundforskningsfond under Grant No.~1041811001.

\appendix

\section{1D cross-disorder correlation functions}

We are interested in computing how spins are correlated across any number of disordered bonds. The object of interest is the correlation function
\eq{}{\langle\sigma_k\sigma_l \rangle - \langle\sigma_k\rangle\langle\sigma_l\rangle} 
where the thermal expectation values are taken in the presence of disorder. We begin with single bond disorder at site $m$, described by the Hamiltonian 
\eq{}{-\beta H = \beta\sum_i\sigma_i\sigma_{i+1} + \beta g_m \sigma_m\sigma_{m+1}}
where we omit subscripts on $\beta$ and $g$ to indicate that we can do this at any particular scale (i.e. after any number of RG iterations). Since the spins can take the values $\pm 1$, the partition function is obtained via the trace
\eq{}{\mathcal Z = {\rm Tr}\,[T^{m-1}\, \bar T_m\, T^{N-m}]}  
where we assume periodic boundary conditions and that there are $N$ sites to sum over. The transfer matrices $T$ and $\bar T_m$ are given by
\eq{transfer}{T = \begin{pmatrix} e^{\beta}&& e^{-\beta}\\ e^{-\beta}&& e^{\beta} \end{pmatrix}~~~\bar T_m = \begin{pmatrix} e^{\beta(1 + g_m)}&& e^{-\beta(1 + g_m)}\\ e^{-\beta(1 + g_m)}&& e^{\beta(1 + g_m)} \end{pmatrix}}
By additivity of the exponents, the matrices $T$ and $\bar T_m$ commute, as would those for any number of disordered bonds at site $j$ denoted $\bar T_j$. Therefore the partition function for a disordered system with $K$ disordered bonds is given by
\eq{}{ \mathcal Z = {\rm Tr}\,[\left(\prod_{j=1}^K \, \bar T_j\right)\, T^{N-K}]}
decimation consists of multiplying matrices pairwise in their original order and rewriting them in the form (\ref{transfer}) with the addition of a free energy that appears as a multiplicative constant, from which one would read off the RG transformations (\ref{it1}) and (\ref{it3}).

Expectation values of the spin operator $\sigma_k$ are obtained by inserting the matrix $\tau$ as the $k$'th matrix in the string of transfer matrices in the trace, with $\tau$ given by
\eq{}{\tau = \begin{pmatrix} 1 &&0\\0&& -1\end{pmatrix} } 
We are interested in calculating correlation functions of spins across a series of disordered bonds at sites $k$ and $l$, where we presume $l>k$. Exploiting the cyclicity of the trace for bounded operators and the fact that $T$ and $\bar T_j$ commute, we can write 
\eq{}{\langle\sigma_k\sigma_l \rangle = \frac{1}{\mathcal{Z}} {\rm Tr}\,[\tau  \left(\prod_{j=1}^K \, \bar T_j\right)\,T^{l-k-K} \tau\, T^{N-l+k}]   }  
where it is convenient to work in the basis where the $\bar T_k$ and $T$ are simultaneously diagonalized, in which case 
\eq{}{\tau \to  \begin{pmatrix} 0 &&1\\1&& 0\end{pmatrix} }
Straightforward calculations then show that $\langle \sigma_k \rangle =0$ and with multiple disordered bonds between the sites $k$ and $l$
\eq{}{\langle\sigma_k\sigma_l \rangle = e^{-\frac{(l-k)}{\xi}}\prod_{j=1}^K\,\frac{{\rm tanh}\,\beta(1 + g_j) }{{\rm tanh}\,\beta}}
where the $g_j$ correspond to the intervening disordered bonds, and where the correlation length $\xi$ one would have obtained in the absence of disorder, which is given by
\eq{}{\xi = -\frac{1}{{\rm log\,(tanh \,\beta)}}}

Note that the fixed points of (\ref{fp}) correspond to the disorder parameter settling on $g = \{-2, 0\}$, so that the factors in the product return $(-1)^{N_{\rm AF}}$, where $N_{\rm AF}$ denotes the number of bonds corresponding to $g=-2$. This has the transparent physical interpretation of preserving the usual correlation up to a sign depending on the number of intermediate twists in the spin chain. The unstable fixed point at $g = -1$ corresponding to a break in the chain contributes a zero, uncorrelating the two spins.

\noindent
\end{document}